# Spatial patterns in the tropical forest reveal connections between negative feedback, aggregation and abundance.

Efrat Seri (efrat083@gmail.com)

Nadav Shnerb (nadav.shnerb@gmail.com)

Department of Physics, Bar-Ilan University Ramat-Gan 52900, Israel

## Abstract

The spatial arrangement of trees in a tropical forest reflects the interplay between aggregating processes, like dispersal limitation, and negative feedback that induces effective repulsion among individuals. Monitoring the variance-mean ratio for conspecific individuals along length-scales, we show that the effect of negative feedback is dominant at short scales, while aggregation characterizes the large-scale patterns. A comparison of different species indicates, surprisingly, that both aggregation and negative feedback scales are related to the overall abundance of the species. This suggests a bottom-up control mechanism, in which the negative feedback dictates the dispersal kernel and the overall abundance.

# Introduction

One of the main characteristics of natural populations, and in particular of sessile species, is their spatial structure. In many cases conspecific individuals are aggregated in space, a phenomena that may be attributed to various mechanisms like dispersal limitation [1,2], positive feedback [3-5] or habitat association. Negative feedback mechanisms, on the other hand, lead to an effective "repulsion" between individuals or clusters [6,7]; the excess competition between same-species individuals may induce self-thinning, and the presence of species-specific parasites or predators around an adult tree may decrease the chance of recruitment in its neighborhood.

Turning from populations to communities, the dynamics of ultra-diverse systems like the tropical forest have attracted a lot of interest, as these systems appear to violate a fundamental assumption of natural selection theory, the competitive exclusion principle [8]. Many mechanistic solutions were suggested to this puzzle [9-15], and almost all of them have to do with some features of these spatial patterns. For example, a competition-colonization tradeoff [15] implies that the better competitors are more clustered in space. Unfortunately, it is quite difficult to relate directly the spatial patterns to the underlying process, since the details of the dynamics (like the recruitment kernel, or the identity of the best competitor) are usually unknown. Still, by pointing out some generic features of the spatial structure of the forest, an analysis may put severe constraints on the suggested models and may serve as a guide for the establishment and refinement of more realistic hypotheses.

A long-standing hypothesis, aimed to explain the apparent access biodiversity of the tropical forest, was suggested by Janzen and Connell [10,11]. Basically, the idea is that host-specific enemies (like pathogens or herbivores) are attracted to an adult tree, making its neighborhood hostile to seeds and seedlings of the same species. Accordingly, conspecific individuals effectively repel each other, implying that inferior species may survive in the forest by filling the gaps between superior competitors.

Recently, the Janzen-Connell hypothesis has gained a renewed popularity and attracted a lot of attention, following a few empirical studies that monitor conspecific negative density dependence and tracked its origin. In particular, a substantial decrease of seed efficiency close to a conspecific adult tree was demonstrated [16,17], and the reduction in the chance of establishment was attributed to the negative effect of soil biota [18]. Moreover, the analysis suggests that the strength of this negative feedback is a good predictor of the commonness/rarity of a species in the tropical forest [17,18]. Similar results (that locally rare species suffer most from the proximity of relatives) were obtained in subtropical [19] and temperate [20] forests.

These new results pose a few interesting questions. The first has to do with the range of the effect. For the Janzen-Connell mechanism to work the negative feedback has to be localized around the adult tree [21]. How such a localized interaction, with a range of, say, a few meters, can affect the community-wide pattern in tree composition [22]? What is the mechanism that allows the negative feedback to dictate features of the forest on much larger scale? One can easily imagine a counter example, where the effect of negative feedback is balanced by another feature. For example, if the seedlings of the "fittest" species (the one that will select out all other species in the absence of negative-feedback mechanisms) cannot establish at all in a radius of 4 meters around an adult tree, this may be the strongest repulsive effect among all species, still the fittest will be one of the most common species in a 500,000m$^2$ forest (like the 50-ha plot in Barro-Colorado Island considered below), with more than 10,000 trees, since its offspring win the competition once they are out of the 4m radius.

This brings us to a second question: the relative strength of this negative feedback mechanism, as oppose to well-known processes that lead to aggregation of conspecific individuals, like dispersal limitation [6]. While the chance of a seed to germinate, or of a seedling establish as an adult, may be smaller close to a conspecific tree, the number of attempts (i.e., the number of seed and seedlings in the vicinity of a reproductive individual) is much larger, and the overall pattern will depend on the interplay between these factors. In particular, very strong negative feedback will leads to a lattice-like forest, while strong aggregating forces yield clumped patterns. In fact, it is well known that a pronounced feature of these spatial patterns is aggregation and clustering [2,23-25], while a direct identification of negative feedback effects from the overall spatial structure of the population has proved itself as quite a difficult task [25]. Given that, one may wonder again about the relevance of conspecific local density dependent to the composition of the community.

In this paper we are trying to shed some light on these problems. Analyzing the spatial patterns that emerge from a few generic models and comparing them to the empirical data, we show that local negative effects and "repulsion" between conspecific individuals indeed dominate the spatial pattern on very short length scales, while aggregation mechanism take over at larger distances.

A more surprising outcome of our analysis emerges when we compare the results for different tree species. It turns out that both the aggregation and the negative feedback are related to the overall abundance of the species in the plot. This phenomenon suggests that the local negative feedback cascades upscales (perhaps by controlling the typical recruitment length) to yield the global pattern. While we cannot suggest a specific mechanistic explanation for these features, we can extract some severe constraints on the possible models of forest dynamics.

This paper is organized as follows. In the methods section we explain the usage of variance-mean ratio and present the results of our analysis for point patterns obtained from a few well-known mechanistic models. The results section is devoted to the analysis of empirical data from the Barro-Colorado island plot (BCI) [26-29], in comparison with the patterns surveyed in the methods, emphasizing the universality of the empirical patterns. Finally, we discuss our result in the general context of variance-mean ratio (Taylor's law) and analyze the apparent insights.

## Methods

The method we implement along this paper is a multiscale analysis of the variance-mean ratio (VMR, also known as index of dispersion, Fano factor). As we shall see, the VMR technique allows for a direct demonstration of both negative feedback and aggregation on different scales.

As long as one deals with a single species, the definition of scale is quite arbitrary: meters, centimeters or the average distance between neighboring trees are all reasonable measurement units. However, in any attempt to compare the statistical features of patterns observed for two (or many) different species, that may have different abundance in the plot, the question of "natural" scale immediately arises. The use of an objective measure, like meters, is natural when the short-range interactions have (almost) nothing to do with the overall abundance, while normalization by a species-specific unit length is required when the length scales are related to the abundance.

In a recent work [30], we have suggested that the second scenario is the relevant one, at least for the tropical forest we have examined. Using other methods of point pattern distribution analysis (nearest neighbor distance distribution, correlation length and cluster statistics), we discovered that spatial patterns of different species obey a universal scaling law once the length is normalized by the typical distance between conspecific trees $\ell_{0i} = \sqrt{A/N_i}$, where $A$ is the area of the plot and $N_i$ is the abundance of the i-th species. Following this result, we implemented here two versions of the VMR analysis: one is based on objective scales, the other utilizes the species specific scale $\ell_{0i}$. As seen in the results section, the VMR analysis seems to support the conclusions of [30], suggesting that spatial patterns of different species are becoming similar once the distances are normalized (for every species) by $\ell_{0i}$.

The index of dispersion is defined, in general, as the ratio between the variance and the mean of a random variable. Here we are looking at a population, i.e., all the individuals of the focal species $i$ in the BCI 50-hactar plot. Covering the plot area by a rectangular mesh of lattice constant $\ell$, and counting the number of individuals in any box, the mean number for the $i$-th species will be $N_i \ell^2 / A$ and the variance depends on the spatial arrangement of the population. The variance mean ratio (VMR) will be a small number if each box contains, more or less, the same number of individuals and will be large when some boxes are almost empty and others are densely occupied so the population is clustered. The degree of clustering may depend on the length scale; accordingly, by plotting the VMR over scales one obtains a summary of the aggregation properties of the system.

Along this paper we present two types of plots. One is a plot of VMR versus the box area $s = \ell^2$, the other is a plot of VMR versus the normalized area $\tilde{s} = (\ell/\ell_{0i})^2$. In the last case length is measured in units of $\ell_{0i}$ and $\tilde{s}$ indicates the mean number of focal species trees in a box.

To wit, let us start with some examples of the VMR-$s$ plots for a few generic point patterns. The simplest case is the Poisson forest, in which trees are spread randomly all over the area. It is well known that, in such a case, the variance is equal to the mean and the VMR is unity, independent the units length used [31].

It is interesting to note that the errors introduced by sampling noise also have similar properties. As long as the noise is uncorrelated among boxes, the sampling is equivalent to the multiplication of the population inside each box by a random number taken from some fixed distribution. In such a case the variance is larger than the mean, but the VMR is still independent of scale. These features of the Poisson forest and the sampling noise are demonstrated in Fig. 1.

Now let us consider the opposite case: a lattice forest. In such a forest a single tree occupies the center of each box of side length $\ell_{0i}$. All boxes with $\ell > \ell_{0i}$ contain the same number of trees, so the variance and the VMR are both zero. For $\ell < \ell_{0i}$ each box is either empty or filled, (moreover, the mesh does not fit the lattice principle axes, and a very weak spatial disorder becomes relevant) and the VMR is unity. Accordingly, the VMR decreases towards zero as $\ell$ increases, as illustrated in Fig. 2.

In a lattice forest the distinction between objective and abundance-dependent scales is clear. Since the only length scale in a lattice forest is $\ell_{0i}$, a plot of the VMR vs. the dimensionless scale $\tilde{s}$ (for different species with different abundance in the same plot) yields a data collapse. On the other hand, if the VMR is plotted vs. $s$ a different curve is obtained for every species and the lines are ordered: since the decay of the VMR from one (Poisson) to zero (lattice) occurs around $\ell \sim \ell_{0i}$, the curve for the most abundant species is the first to deviate from the Poisson value, then the second and so on.

In general, the decay of the VMR on increasing length scales is a hallmark of negative feedback: when individuals repel each other, either via competition or by attracting hostile parasites, the spatial structure becomes lattice-like and the VMR decays at large distances. The opposite happens when individuals are aggregated. Again, at short length scales, the VMR must be Poisson-like, but for larger scales the variance increases faster than the mean, leading to an increasing VMR - $s$ line [32]. Figure 3 exemplifies this property for a fractal forest, a structure suggested as a model for the BCI by Ostling et. al. [33]. In such a fractal forest the typical distance between neighboring conspecific trees is a fixed number $\ell_{fractal}$, but $\ell_{0i}$ depends on the overall species abundance. As a result, the VMR-$s$ plot shows data collapse for different species, while in the VMR-$\tilde{s}$ the curve deviates from the Poisson value when $\ell_{fractal} \sim \ell$, so the curve of the rarest species starts to increase earlier. These characteristics of the VMR do not depend on any special features of a specific fractal structure (like the random Cantor set used here); they do hold also for other fractals.

A more mechanistic model of spatial aggregation is suggested by models that take into account the details of the underlying birth-death process. In particular, the neutral theory of biodiversity suggests a simple framework, in which at every elementary timestep one tree is chosen to die and is replaced by the offspring of another tree, chosen at random from the neighborhood of the dead individual [2]. Starting from the founder of the focal species, one can run a simulation of forest dynamics until the desired abundance is reached. This simulation is quite simple, since the assumption of neutrality saves the need to distinguish between non-focal species. Accordingly, the only parameter that affects the spatial structure is the recruitment kernel, i.e., the chance that an offspring from a tree which is located at a distance r from the dead tree will capture the empty slot.

When the recruitment kernel is infinite (the chance of any tree to replace any other is distance-independent) the forest is Poissonian. Here we consider two generic kernels: the mixed local-global kernel (MLGK, the reproducing tree is chosen, with probability μ, at random from the whole forest and with probability 1-μ from a 2-meter neighborhood of the dead tree. Clearly μ=1 is the Poissonian limit) and the Cauchy kernel where the reproducing tree is chosen with probability $P(r)$ that corresponds to the (fat-tailed) Cauchy distribution, $P(r) \sim 1/\left[1+(r/\ell_{rec})^2\right]$. In a previous work [2] we have tried to fit the cluster statistics of the BCI using these two kernels, and showed that the Cauchy kernel fits much better than the MLGK. Here we apply the VMR analysis to a simulated forest with these kernels, where the parameters used are those that gave the best fit to the BCI in [2].

Under neutral dynamics there are two length scales associated with each species: $\ell_{0i}$ that depends on the abundance and the scale associated with the recruitment kernel, $\ell_{rec}$. These scales are independent, and one should expect that the VMR shows substantial deviations from Poisson once $\ell > \ell_{rec}$. Accordingly, the VMR-$\tilde{s}$ plot shows a clear order (the rare species curves raise first), while in the VMR-$s$ diagram the data collapse (since we have taken the same $\ell_{rec}$ for all species).

The MLGK kernel yields a spatial structure that resembles the patterns observed for a Cox process (where centers are chosen at random and then trees are spread, again at random, in the neighborhood of each center [34]); dispersal limitations induce clustering only within each local patch, but the large scale structure of the forest is Poissonian. Therefore, the VMR curve "bends over" towards the Poisson limit for $\ell \gg \ell_{rec}$ (since our forest is finite the large $\ell$ behavior is noisy, but the trend is clear), yielding a hump-shaped graph (Fig. 4). For a Cauchy forest the length scale associated with the average recruitment radius is infinite and the VMR-area curves are monotonic. In fact, one can fit these curves quite nicely with the expression $a + b\tilde{s}^z$ [32], where for all species $a \approx 1$ (in the short scale Poissonian regime) and the values of the exponent $z \approx 0.5$ are also abundance independent, reflecting only the features of the recruitment kernel (Fig. 5).

The results for all these mechanistic models are summarized in Table 1. Evidently, there are two essential qualitative characteristics for any VMR graph. The first is its behavior along scales, where an increase means aggregation, a decrease implies negative feedback and repulsion, and a constant value suggests a Poisson distribution. The second characteristic emerges when the VMR's of species with different abundance is compared. If the typical distance between neighboring trees is $\ell_{0i}$ the VMR-$\tilde{s}$ diagram shows a data collapse, and the frequent species are the first to leave the Poisson regime in the VMR-$s$ plot. On the other hand, when the typical scale between neighboring trees is independent of $\ell_{0i}$ (e.g., the recruitment kernel) the collapse is manifested only in the VMR-$s$ plot, and the rare species are the first to show substantial deviations from the Poisson limit in the VMR-$\tilde{s}$ diagram.

# Results

Given the insights gained from the analysis of VMR plots for these generic models, we are in position to apply the same analysis to a real community of trees in the tropical forest. A-priory, one should not expect a perfect data collapse as those observed for the models, since the natural population dynamics is affected by all kinds of stochastic forces, and perhaps the assumption that trees of different species have the same spatial dynamics (e.g., the same recruitment kernel) is, in the best case, only a crude approximation. Still it is interesting to examine the empirical results and to find out if they show any kind of qualitative similarity to the models considered so far.

Figure 6 show the VMR-$\tilde{s}$ and the VMR-$s$ plots for all the dbh>1cm individuals of the 43 $N_i > 1000$ species in the BCI forest. These results are ambiguous: on the one hand, the gross features of the plot resemble those of the neutral Cauchy forest: a Poisson region for small length scales followed by a power-law growth at large scales (the Fractal forest has also these feature, but the data from the BCI do not show the spatial "macro-gaps " like those observed in Fig. 3). On the other hand, while none of the panels shows an impressive data collapse, the VMR-$\tilde{s}$ curves appears to fall in a much narrower region, suggesting that the more relevant length-scale is $\ell_{0i}$.

The idea of a forest with Cauchy-like recruitment kernel appears to be plausible. It agrees (as mentioned above) with our detailed analysis of the spatial patterns for the most abundant species [2] and with the fractal analysis of [1]. For all species $a \approx 1$ indicating that the sampling errors are small. The exponent $z$ has values for all the species [see insets of Figs 5(a) and 6(a)] suggesting that the long-distance properties of the spatial dynamics are governed by fat tailed processes. To explain the absence of a data collapse in the VMR-s plot one may propose that different species has different typical recruitment scale $\ell_{rec}$. Nevertheless, the quasi-collapse in the VMR-$\tilde{s}$ plot suggests a more radical insight: that the recruitment kernel of a species is proportional to $\ell_{0i}$, i.e., that there are correlations between the typical distance between a tree and its offspring (a "local" characteristics of the dynamics) and the overall abundance of this species in the forest – a "global" feature. This conclusion is in agreement with another study that our recent study [30], supporting this "glocality" by implementing other point-pattern analyses.

An interesting finding was discovered when we considered the short-range region of Fig. 6 in more detail. Although the VMR curves at short scales appear to stick to one as expected in the Poisson regime, a zoom into the small *s* region reveals a weak, but pronounced, *repulsion (anti-correlation)* between conspecific trees, with a typical *decline* of VMR below one as observed for a lattice forest (Fig. 7). Although this submetric deep is weak and very noisy, it appears in many species and, unless it reflects an artifact of the data collection procedure, seems to indicate repulsion. Note that the weakness of the negative feedback signal does not imply that the effect itself is weak, since it competes with the aggregation mechanism (dispersal limitation, say) over all scales.

Amazingly, the negative feedback mechanism demonstrated in fig. 7, seems also to be related to the global scale $\ell_{0i}$. First, the lines in the repulsion zone appear to follow a species-independent curve in the VMR-$\tilde{s}$ plot, while there is no such a feature in the VMR-$s$. Second, only in VMR-$\tilde{s}$ plot the strength of the repulsion is (negatively) correlated with the height of the VMR in the attractive regime, e.g., at $\ell/\ell_{0i}=1$. This feature is depicted in panels (c) and (d) of Fig. 7, where the attraction at a fixed rescaled distance is plotted against the strength of repulsion (measured by the inverse distance between the minimal value of VMR for a certain species and the Poisson level at VMR=1). The weaker is the repulsion, the higher is the VMR at $\tilde{s} = \ell/\ell_{0i} = 1$. This correlation does not appear in the VMR-$s$ plot.

The simplest (although quite radical) interpretation of our results is that the negative feedback controls both the aggregation and the overall abundance of a species in the forest; hence all these patterns are related to the same scale.  A possible explanation is that the evolutionary development of dispersal strategies may be governed by the negative feedback so the recruitment kernel increases with the strength of the local negative density dependence, otherwise the loss of seeds in the prohibited zone will lower the species' fitness and will lead to extinction.

## Discussion

The scaling of the variance with the mean is known as an interesting statistical parameter used to characterize the fluctuations in a system. In particular, the index of dispersion (ID) is widely used in the analysis of ecological point patterns [36]. The usage of this statistic as an indication for Poisson distribution was criticized by many authors, since the VMR is one for a whole set of non-Poissonian distributions (the so called 'unicornian' distributions [37]) but these distributions, in general, do not keep this unicornian property under spatial rescaling (i.e., they are not "Tweedy" [31]). The variation of the VMR (Fano factor) over scales is a very common analytic tool in other branches of science and in particular in the analysis of neural spike trains, where the deviations from the Poissonian limit indicate either aggregation or repulsion [35].

One of the popular methods used in ecology to characterize spatial clumpiness is the exponent of the variance-mean curve, or its slope when plotted on a double-logarithmic scale. Comparing the variance-mean relationships of many censuses that involved quadrants of different size, Taylor [38] decided that the variance scales like a power of the mean, with an exponent $b$ that, in most cases, falls in the region $1 < b < 2$. Clearly, the variance-mean slope and the VMR considered here are related and $b=z+1$. As showed here, Taylor's law cannot hold on small scales (at least for scales that are shorter than the typical distance between two individuals) where the fluctuation statistics must become Poissonian. This observation is compatible with other works that criticized Taylor's conjecture [32].

However, as already suggested by Nedler [32], in the intermediate scales we have found, indeed, a reasonable fit to the variance-mean relation $VMR = a + b\tilde{s}^z$ over 2-3 decades. The exponent $z$ appears to be in a narrow range of parameters for almost all the species considered here, once the length scale is normalized by $\ell_{0i}$, i.e, once the VMR is plotted against the mean ($\tilde{s}$). This observation suggests that all species are subject to the same kind of spatial dynamics, where the only difference is the basic length scale associated with the overall abundance of the species in the forest $\ell_{0i}$. It also agrees with our new work [30], where we show that a few basic characteristics of the spatial structure become similar for all the species under species-specific normalization of the length scale by $\ell_{0i}$.

The plots of the VMR along scales, presented here, allow us to point out a possible underlying mechanism. The weak decrease of the VMR below its Poisson limit is an evidence for a short-range repulsion between conspecific trees. Accordingly, the VMR at any scale reflects the interplay between short-range repulsion and the intermediate range aggregating mechanisms.

As seen from the numerical simulations, when the dynamics has no repulsion the VMR at $\tilde{s}=1$ is decreasing as the abundance is increasing. The fact that this relation is destroyed by the repulsive forces implies that the strength of repulsion is correlated with $\ell_{0i}$, i.e., with the overall abundance of the species in the forest. The "universal curve" that characterizes the repulsion in the VMR-$\tilde{s}$ plot (Fig. 7), and the absence of correlations between the repulsion and the VMR at the non-normalized scale $s=10m$, also point toward this hypothesis. If true, these findings appear to be consistent with the conclusion of [17,18,20] and others, and to suggest that the repulsive interaction affects the overall abundance since it governs the process that leads to aggregation.

The suggested correlation between negative feedback and abundance seems to put severe restrictions on any proposed underlying mechanistic model. It is clear that in a model where species differ by their competitive ability (i.e., in a site which is far away from any conspecific adult, the seeds/seedlings of species A have a better chance to establish than the seeds/seedlings of species B) this feature will dictate the large-scale abundance and will destroy the correlation between negative feedback and $\ell_{0i}$. Accordingly, a minimal model that will preserve the features demonstrated here is a generalized version of Hubbell's neutral model of biodiversity [9], in which every species has its own "typical distance" that dictates both the negative feedback and the dispersal kernel, but otherwise all individuals are equal, i.e., the chance of seedlings to capture a site is independent of species identity once all these seedlings are out of the negative feedback zone. Such a model may settle the apparent contradiction between Hubbell's version of the neutral theory, in which species identity has nothing to do with its abundance, and the long distance correlations between the abundance of trees that belong to the same family pointed out recently by [39]. However, it is not trivial that the nice features of the neutral dynamics, like coexistence and realistic species-abundance distributions in a mainland-island setup, are preserved when a species-specific length scale is introduced. We hope to consider this problem in a future work.

**Acknowledgements**: We thank David Kessler, Lewi Stone, Joe Wright and Richard Condit for helpful discussions and comments. This work was supported by the Israeli Ministry of Science and Technology TASHTIOT program and by the Israeli Science Foundation grant no. 454/11 and BIKURA grant no. 1026/11. The BCI forest dynamics research project was made possible by National Science Foundation grants to Stephen P. Hubbell: DEB-0640386, DEB-0425651, DEB-0346488, DEB-0129874, DEB-00753102, DEB-9909347, DEB-9615226, DEB-9615226, DEB-9405933, DEB-9221033, DEB-9100058, DEB-8906869, DEB-8605042, DEB-8206992, DEB-7922197, support from the Center for Tropical Forest Science, the Smithsonian Tropical Research Institute, the John D. and Catherine T. MacArthur Foundation, the Mellon Foundation, the Small World Institute Fund, and numerous private individuals, and through the hard work of over 100 people from 10 countries over the past two decades. The plot project is part the Center for Tropical Forest Science, a global network of large-scale demographic tree plots.

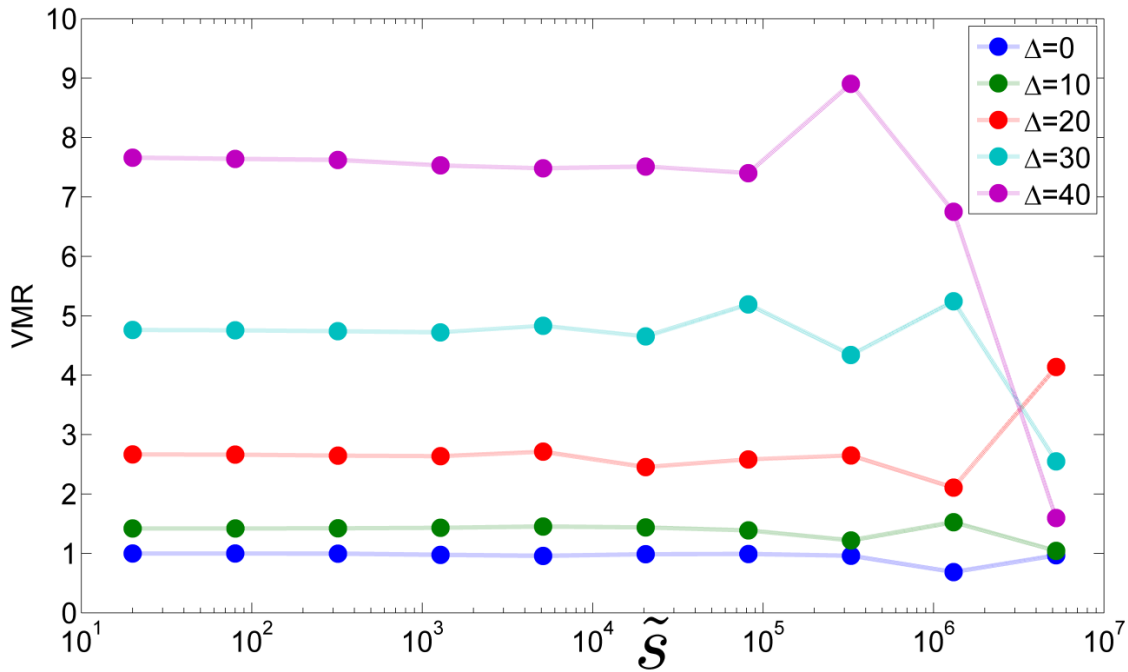

**Figure 1.** VMR versus $\tilde{s}$ for a Poisson forest, with different levels of sampling error $\Delta$. When there is no sampling error ($\Delta = 0$) the VMR equals unity and is independent of scale as expected. The graphs for $\Delta > 0$ were obtained for different strength of the sampling error. In the simulations, an elementary box of arbitrary area was defined, and for each box j a random number $\eta_j = \Delta \cdot [-0.5, 0.5]$ is assigned. The number of trees at any box was then picked at random from a Poisson distribution with an average $20 + \eta_j$, so for $\Delta = 40$ one obtains the maximal sampling noise. Clearly, sampling errors increase the size of the fluctuations, but (as long as the error is spatially uncorrelated) the VMR curve is still $\tilde{s}$ independent.

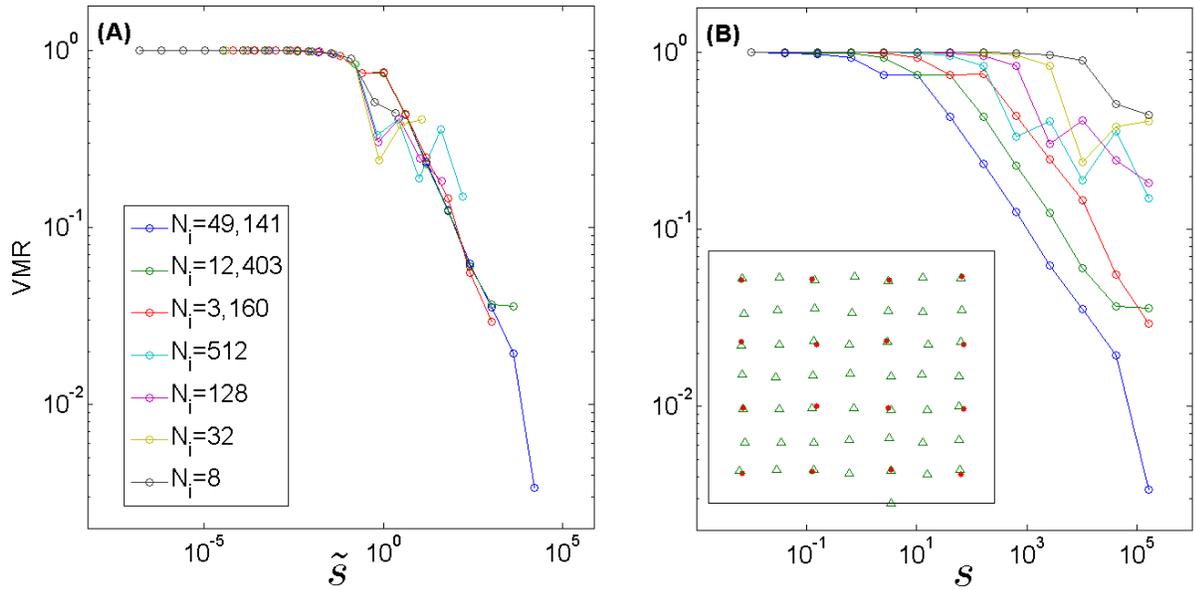

**Figure 2.** The lattice forest. VMR vs. $\tilde{s}$ (A) and vs. $s$ (B). For every species of $N_i$ trees, a square lattice with unit length $\ell_{0i}$ admits $N_i$ vertices in a forest of area A. In the simulated forest a single tree is located at random within a fixed distance (much smaller than the lattice constant) around every vertex. Example of a two species (frequent – green triangles, rare – red points) lattice forest is given in the inset of panel (B). The VMR at short length scales is still unity, due to the weak noise, but decreases to zero at large scales, when every box contains the same number of trees. The VMR line shows substantial deviations from the Poisson limit when the average number of trees inside a box is one. Accordingly, the frequent species VMR is the first to decay (panel (B)). On the other hand, when the VMR is plotted against $\tilde{s}$ the curves collapse. The results are shown here for species with different $N_i$-s, to reflect the range of abundance one finds in the BCI. The mean of 10 curves are presented for every frequency class, where different colors correspond to different $N_i$-s. The numbers $N_i$ and the size of the plot (a rectangle of 500X1000 meters) were chosen at order to mimic the abundance classes and the area of the BCI forest, both here and in the following figures.

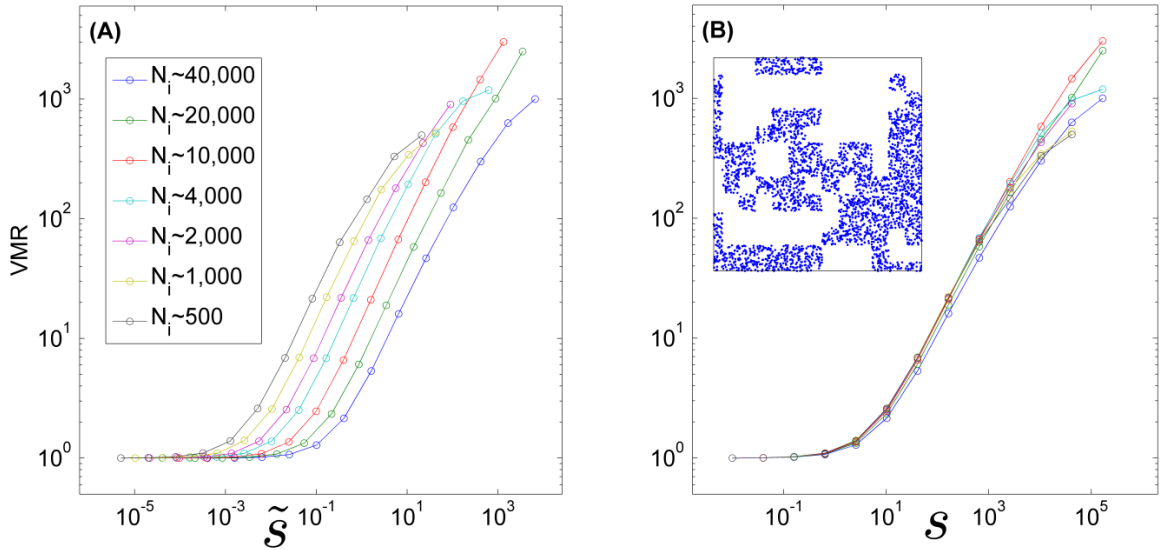

**Figure 3**. The fractal forest: VMR for a random Cantor set forest vs. $\tilde{s}$ (A) and vs. $s$ (B). We have run the algorithm suggested in Ostling et al., starting with a 2x2 array. Each cell is chosen to be empty with probability P or to be "active" with probability 1-P. An active cell is then divided into 4 equal squares and the process is iterated. The process (using P=0.75) stopped when the forest reaches a size of 256x256 cells. A single realization of the fractal is shown in the inset of panel (B). The area of any elementary box is defined to be 16m$^2$ and $n$=40 trees were located at random within each active box. Implementing a few realizations of the same algorithm, we were able (due to the randomness of the process) to generate a few sets of focal species trees, sets that have the same fractal structure but different abundance. For the analysis presented below we have grouped together realizations that have, more or less, $N_i$ trees and show the mean of them. As expected, the VMR deviates upward from unity at intermediate length scales. Since the minimal distance between neighboring conspecific trees is independent of the abundance of a species, the curves collapse in the VMR-$s$ plot, while in the VMR-$\tilde{s}$ plot the rarer is the species, the lower is the point in which its curve begins to bend upward.

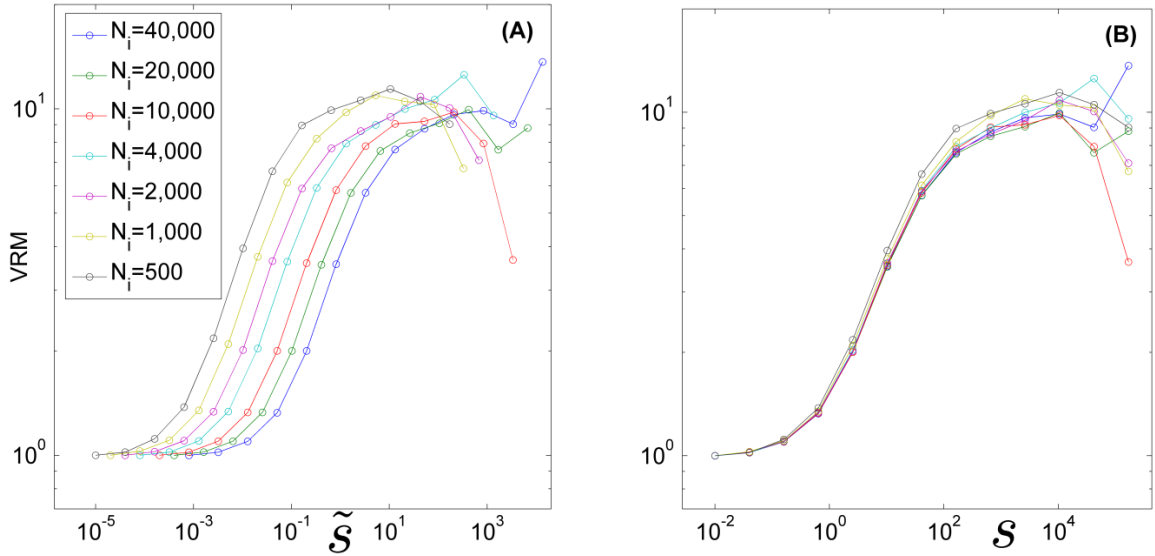

**Figure 4.** A neutral forest with local-global recruitment kernel. VMR vs. $\tilde{s}$ (A) and vs. $s$ (B). The dynamics is neutral, and the LG recruitment kernel is applied with $\mu = 0.1$, which is the best fit parameter to the cluster size statistics [2]. To imitate the BCI forest the simulations run over 500x1000 m rectangle, with a neutral dynamics used in [2], until the abundance reaches $N_i$ trees. The mean of ten iterations for every value of $N_i$ is shown. The Poisson region is evident on short scales, and clustering manifests itself on the intermediate scales. On larger length scales the curves must return to the Poisson limit, as clusters locations are uncorrelated. Here (for simulations on the scale of the BCI forest, and with the relevant parameters) one can see only the early onset of the decrease. On large scales the curve should return to unity.

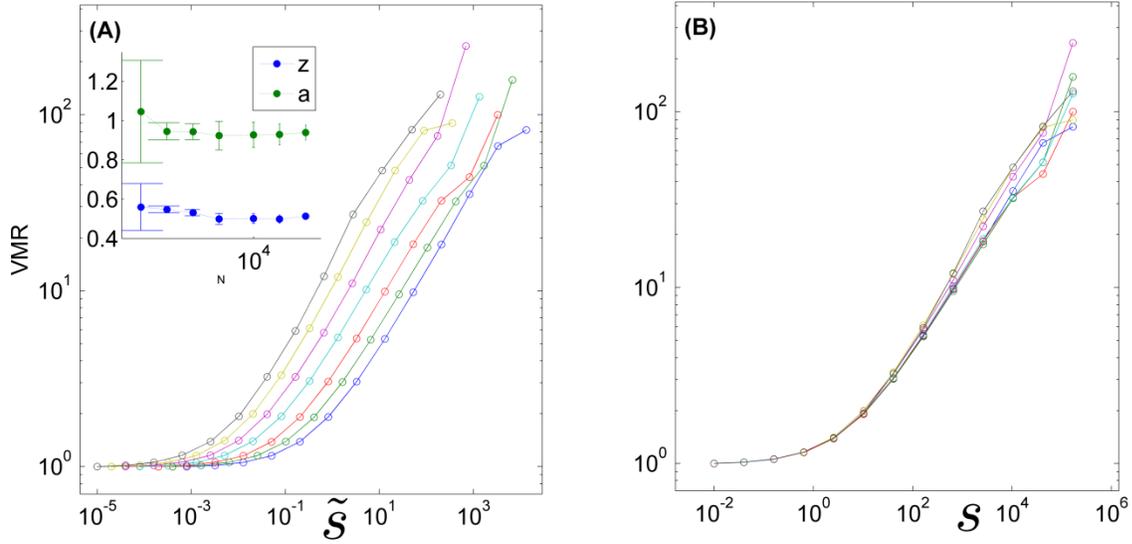

**Figure 5.** The neutral-Cauchy forest. VMR vs. $\tilde{s}$ (A) and vs. $s$ (B). The dynamic is the same as in fig. 4 but here Cauchy recruitment kernel is applied with $\gamma = 20$, which is the parameter that yields the best fit to the cluster size statistics [2]. One may see the Poisson region in short scales, and the intermediate scales of clustering, but there is no return to the Poisson limit on large scales. The inset in (A) shows the parameters of the VMR fit to $a + b\tilde{s}^z$. The fit for specific species is for the average of 10 iterations and for the 11 first points (the last points are too noisy). $R^2 > 0.999$ for all the fits shown in this figure.

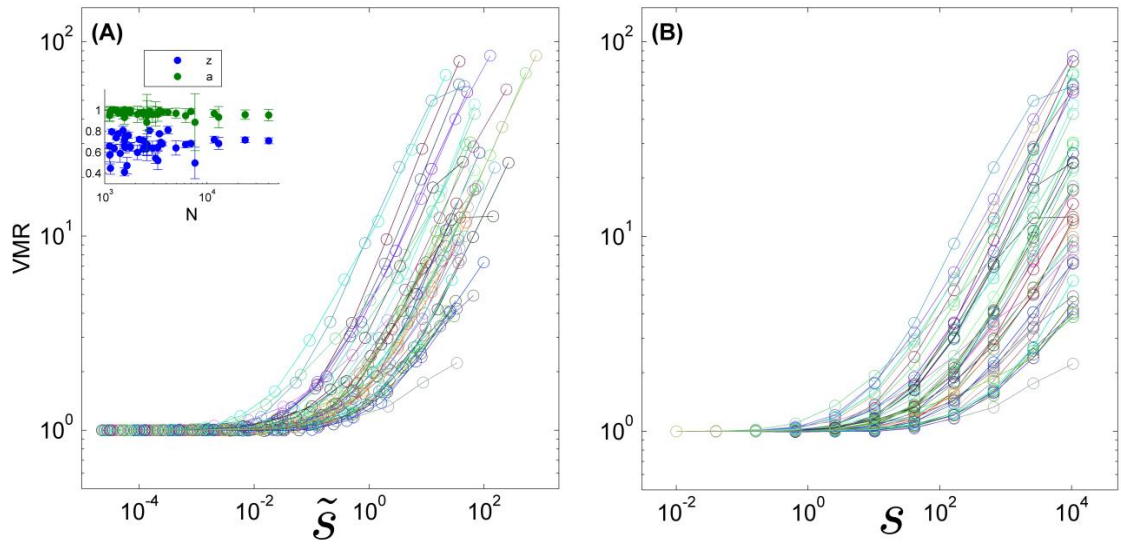

Figure 6. VMR for the real data from the BCI plot. The VMR vs. $\tilde{s}$ (A) and vs. $s$ (B) is shown for all the 43 species with abundance > 1000 trees. The inset in (A) shows the parameters of the fit to $a+b\tilde{s}^z$, here in most of the curves $R^2>0.99$ (except two for which $R^2>0.96$).

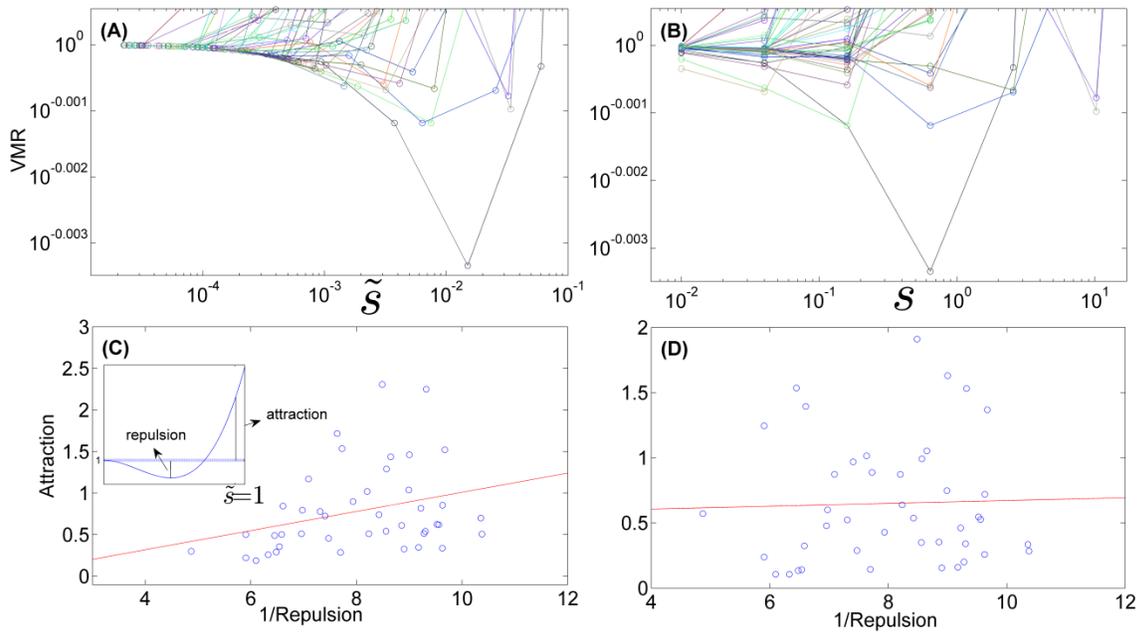

Figure 7. The deep at short length scales. Panels (A-B) show the same datasets presented in fig.6, zooming into the short length scale. In panels (C-D) The attraction parameter (VMR at $\tilde{s}=1$) is plotted against the inverse of the repulsion parameter (VMR at the deep maximum, see the inset of C). The red lines are the linear fit, showing almost no correlation in (D) (Pearson correlation coefficient is 0.03 and the p-value is 0.8) but quite pronounced correlation in (C) (Pearson 0.3, p-value less than 0.05).

**Table 1**

| Length scale that governs the typical distance between neighboring trees | Model(s) | VMR | VMR-$s$ for different N-s | VMR-$\tilde{s}$ for different N-s |
|---|---|---|---|---|
| $\ell_{0i}$ | Poisson | Stays fixed | Data Collapse | Data Collapse |
| $\ell_{0i}$ | Lattice | Decreases to zero | No collapse, abundant species are the first to descend. | Data Collapse |
| Recruitment kernel $\ell_{rec}$ | MGLK | Hump-shaped | Data Collapse | No collapse, rare species are the first to ascend. |
| Recruitment kernel $\ell_{rec}$ | Cauchy | Increases | Data Collapse | No collapse, rare species are the first to ascend. |
| Fractal unit length | Random Cantor set | Increases | Data Collapse | No collapse, rare species are the first to ascend. |

28